\documentclass[pdflatex,sn-mathphys-num,iicol]{sn-jnl}

\usepackage{graphicx}
\usepackage{amsmath,amssymb,amsfonts}
\usepackage{amsthm}
\usepackage{physics}        % optional, if you like \dv, \ket, etc.
\usepackage{hyperref}       % recommended
\usepackage{cleveref}       % recommended
\usepackage{slashed}
\usepackage{physics}
\usepackage{mathrsfs}
\usepackage{soul}
\usepackage[normalem]{ulem}
\newcommand{\lambdabar}{\ensuremath{\mathchar'26\mkern-9mu \lambda}}

\theoremstyle{plain}

\theoremstyle{definition}

\theoremstyle{remark}

\begin{document}

\title[short title]{Will we ever quantize the centers of mass of heavy systems?\\ A case for a Heisenberg cut in quantum mechanics.}

% -----------------------
% Authors and affiliations
% -----------------------

\author[1]{\fnm{Gabriel H. S.} \sur{Aguiar}}
\email{ghs.aguiar@unesp.br}

\author*[1]{\fnm{George E. A.} \sur{Matsas}}
\email{george.matsas@unesp.br}

\affil[1]{\orgdiv{Instituto de F{\'i}sica Te{\'o}rica}, \orgname{Universidade Estadual Paulista}, \orgaddress{\city{S{\~a}o Paulo}, \state{S{\~a}o Paulo}, \country{Brazil}}}

% -----------------------
% Abstract and keywords
% -----------------------

\abstract{
The quantum orthodoxy attests that all dynamical degrees of freedom should be quantized, including those associated with the centers of mass of composite/complex systems, such as protons/pebbles. By assuming that the centers of mass of protons/pebbles should be quantized, one implies the existence of a corresponding Fock space, since, according to the present lore, quantum mechanics is a consequence of quantum field theory. Despite the fabulous success of quantum mechanics, it is unreasonable to assume the existence of annihilation and creation operators for pebbles. Fortunately, there are strong reasons to doubt that wave mechanics can describe the centers of mass of systems at or above the Planck scale, thereby jeopardizing the construction of the corresponding Fock space. We argue that isolated (free) systems with masses exceeding the Planck mass would have their centers of mass governed by classical (rather than quantum) mechanics, despite harboring macroscopic quantum phenomena as observed in the laboratory. Here, we briefly revisit (i)~the arguments for the need for a Heisenberg cut delimiting the boundary between the quantum and classical realms and (ii)~the kind of new physics expected at (the uncharted region of) the Heisenberg cut.}


\maketitle

% -----------------------
% Main text\wp\wp\partial
% -----------------------

%%%%%%%%%%%%%%%%%%%%%%%%%%%%%%%%%%%%
\section{Introduction}
%%%%%%%%%%%%%%%%%%%%%%%%%%%%%%%%%%%%

The stupendous success of quantum mechanics~(QM) in describing elementary particles and the fact that elementary particles are the constitutive blocks of macroscopic bodies have led to the widespread belief that QM would suffice to account for the macroscopic world. As reasonable as it may seem, the assumption that QM is universally valid ultimately proves problematic. Quantum particles are a concept derived from quantum field theory, whose Fock space is constructed based on wave mechanics. From accepting that wave mechanics describes the centers of mass (c.m.) of composite bodies as massive as pebbles, it follows that they would have a corresponding Fock space, annihilation and creation operators, and so on. This is not only strange but also physically unfounded. The reason can be summed up in the fact that {\em wave mechanics presupposes an underlying spacetime~$(\mathscr{M}, g_{ab})$, defined by a smooth manifold~$\mathscr{M}$ endowed with a Lorentzian metric~$g_{ab}$, and there are cogent reasons to doubt the equations of wave mechanics for masses at the Planck scale, $M_\text{P} \sim 10^{- 5}~{\rm g}$ (and above it).} Before proceeding, we shall pause to discuss the system of units adopted here. Our conclusions do not depend on it, but the simplification it yields pays off. 

%%%%%%%%%%%%%%%%%%%%%%%%%%%%%%%%%%%%%%%%%%%%%%%%%%%%%%%%%%
\section{Essential system of units}
%%%%%%%%%%%%%%%%%%%%%%%%%%%%%%%%%%%%%%%%%%%%%%%%%%%%%%%%%%

We will work with units in which 
$$
    k_\text{B} = k_\text{e} = G = c = 1,
$$ 
thereby eliminating all irrelevant conversion factors, in which case all observables are expressed in time units. (Note that~$\hbar$ is not in the above list!) The message here is clear: {\em the same and sole apparatus with which relativistic spacetimes~$(\mathscr{M}, g_{ab})$ must come equipped to be tested, namely, bona fide clocks, is enough to measure all observables defined over them.} We direct the reader to Ref.~\cite{MPSV2024} for a comprehensive discussion, but let us underscore the pivotal points that support this conclusion. 

First, we recall that the International System~(SI) was established in 1960, based on the MKS system; so, $k_\text{B}$ and~$k_{e}$ are mere conversion factors from MKS into kelvin and ampere units of the SI, namely, from~${\rm kg} \cdot {\rm m}^2 \cdot{\rm s}^{- 2}$ to~${\rm K}$ and from~${\rm kg}^{1 / 2} \cdot {\rm m}^{3 / 2} \cdot{\rm s}^{- 2}$ to~${\rm A}$, respectively. It rests, thus, to see that $G$ and $c$ (in relativistic spacetimes) are also conversion factors.

The kg was introduced during the French Revolution (for commercial reasons) about 70 years after Newton's death. Newton did not require a mass standard to make his predictions, nor did anyone else. This common wisdom has faded over time. About 150 years ago, {\em e.g.}, we can see Maxwell writing in the preliminaries of his masterpiece {\em A Treatise on Electricity and Magnetism:} ``If, as in the astronomical system, the unit of mass is defined with respect to its attractive power, the dimensions of [mass] are $\text{[length]}^3/\text{[time]}^2$'', meaning that it can be measured with rulers and clocks. To determine the mass~$m$ of a body, one must measure the acceleration~$a$ with which a test particle drops towards the body when it is at an arbitrarily large distance~$L$: $m = a \times L^2$. The conversion factor between $\text{m}^3/\text{s}^2$ and $\text{kg}$ was determined 80 years after the introduction of the kg. (Before it, mass units used in science and commerce did not communicate with one another.) The conversion factor is the well-known gravitational constant, $G$. In the absence of~$G$, all observables would be expressed in space and time units. Hence, following the adage that it never hurts to find oneself aligned with Newton, we set $G = 1$, thereby vanishing the kg standard from the MKS system.
\begin{figure}[!ht]
    \begin{center}
    \includegraphics[scale = 0.5]{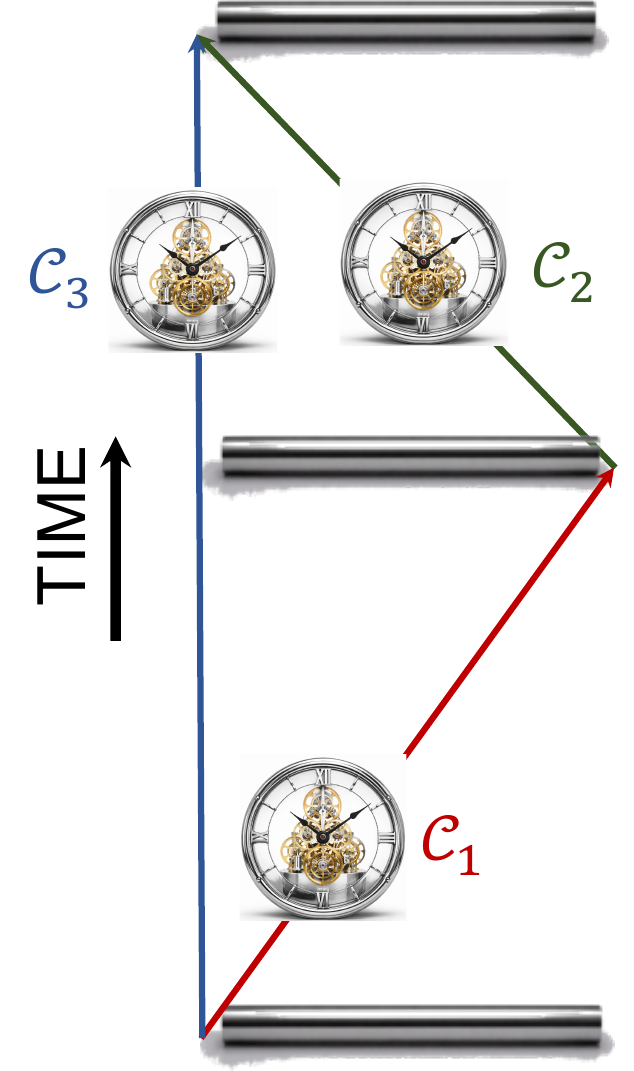}
    \end{center}
    \caption{The figure shows clock~${\cal C}_1$ moving from the left to the right along a small rod and measuring the proper time~$\tau_1$ of its one-way trip, clock~${\cal C}_2$ measuring the return-trip proper time~$\tau_2$, and clock~${\cal C}_3$ measuring the time interval between the departure of~${\cal C}_1$ and returning of~${\cal C}_2$. (Figure generated with AI help.)}
    \label{rod}
\end{figure}

So far, so good, but what about the speed of light~$c$? Relativistic spacetimes are manifolds endowed with Lorentzian metrics. To be defined and tested, they must be equipped with bona fide clocks. Bona fide clocks ascribe the same real number to any pair of arbitrarily close events they visit, irrespective of their state of motion and past histories. Not only all spacetime observables, such as the size of black holes, but also all physical observables defined in relativistic spacetimes may be expressed in terms of time units given by bona fide clocks~\cite{MPSV2024}. Perhaps the most straightforward way to convince oneself of it is through a protocol devised by Unruh. To measure, say, the length~$D$ of a small rod in a relativistic spacetime, three clocks suffice. The first clock~$\mathcal{C}_1$ flies from the left to the right end of the rod, recording the time interval~$\tau_1$ of its trip. Immediately afterward, a second clock~$\mathcal{C}_2$ is sent back, recording the return time interval~$\tau_2$. Meanwhile, a clock~$\mathcal{C}_3$ records the time interval~$\tau_3$ that elapses from the departure of~$\mathcal{C}_1$ to the return of~$\mathcal{C}_2$. In relativistic spacetimes, the rod's length in time units (related, {\em e.g.}, with light-seconds) will read~\cite{MPSV2024}
\begin{equation}
    D = \left[(\tau_1^2 + \tau_2^2 - \tau_3^2)^2 - 4 \tau_1^2 \tau_2^2 \right]^{1 / 2} / (2 \tau_3),
    \label{UE}
\end{equation}
irrespective of how quickly $\mathcal{C}_1$ and~$\mathcal{C}_2$ move back and forth. (Note that when clocks $\mathcal{C}_1$ and $\mathcal{C}_2$ approach the speed of light, $\tau_1=\tau_2\to 0$, we get $D \to \tau_3/2$, as it should be.) It is clear, hence, that distances can be measured only with clocks {\em in relativistic spacetimes}. In the case that all physical quantities are measured with clocks, all velocities are dimensionless numbers, expressing how close something moves with respect to the causality (light) cone, $c = 1$. The speed of light~$c$ acquires dimensional units only when it serves as a conversion factor from time units to a human-made distance unit, such as the meter. It is worthwhile to note that in Galilei spacetime $\tau_3 = \tau_1 + \tau_2$, driving the right-hand side of Eq.~\eqref{UE} to vanish identically. The fact that Eq.~\eqref{UE} is invalid in Galilei spacetime simply reflects the necessity of independent standards for space and time. In Galilei spacetime, $c$, as any other speed, is a legitimate physical observable.  

To finish this section, let us comment that in units where~$k_\text{B} = k_\text{e} = G = c = 1$, $\hbar \approx 2.9 \times 10^{- 87}~{\rm s}^2$. We stress that $\hbar$ is a physical observable that gives the scale of the elementary angular momentum and should not be mixed up with man-made conversion factors such as~$k_\text{B}$, $k_\text{e}$, $G$, and~$c$ (in relativistic spacetimes). 

%%%%%%%%%%%%%%%%%%%%%%%%%%%%%%%%%%%%%%%%%%%%%%%%%%%%%%%%%%
\section{Limits for spacetime theories}
%%%%%%%%%%%%%%%%%%%%%%%%%%%%%%%%%%%%%%%%%%%%%%%%%%%%%%%%%%

As mentioned above, bona fide clocks must assign the same real number to any pair of arbitrarily close events, irrespective of their state of motion and past histories, according to the theory of relativistic spacetimes. However, one thing is what a theory requires to make sense; something completely different is whether nature can provide it. Thus, we must scrutinize the extent to which bona fide clocks exist in nature. 

To comply with the demand above, bona fide clocks should (i)~be pointlike and (ii)~possess a well-defined worldline. As dynamic systems, clockworks are expected to act in accordance with quantum mechanics, impeding bona fide clocks from perfectly complying with~(i) and~(ii). The best trade-off in practice for bona fide clocks is to regard them as composed of a quantum clockwork encapsulated in a (as small as possible) classical container. The question we must now address is whether these bona fide clocks can measure Planck-scale intervals in space and time. 

Let us reconsider the enterprise of measuring the length of a rod, now taking into account physical clocks subject to quantum-mechanical constraints. For this purpose, one can either use the three-clock protocol described above or, equivalently, measure (half of) how long a light ray takes to make a round-trip along the rod. For the sake of argument, let us begin by considering a spherical clock with a mass~$M_\text{clock}$. To avoid collapsing into a black hole, the clock radius must exceed its Schwarzschild radius
\begin{equation}
    R_\text{S} \equiv 2 M_\text{clock} > 2 M_\text{cw}, 
    \label{line1}
\end{equation}
where the last inequality states that the clock mass, $M_\text{clock}$, must exceed that of the clockwork, $M_\text{cw}$, since it must also account for the container mass. 

Next, we use the standard quantum-mechanical result (see, {\em e.g.}, Ref.~\cite{2010-SW}) that in order to acquire a time precision of~$\delta t$, the clockwork is required an energy not smaller than $\hbar / (2 \delta t)$:
\begin{equation}
    M_\text{cw} \geq \hbar / (2 \delta t) > \hbar / (2 D), 
    \label{line2}
\end{equation}
where the last inequality recalls that~$\delta t$ must be smaller than the rod's length~$D$, {\em e.g.}, to measure the Earth-Moon distance, of about 1~(light-) s, one needs clocks with precision~$\delta t < 1~{\rm s}$. Combining Eqs.~\eqref{line1} and~\eqref{line2}, we obtain
\begin{equation}
    R_\text{S} > \hbar / D. 
    \label{line3}
\end{equation}
Now, let us assume that our rod is smaller than the Planck length~$L_\text{P} \equiv \hbar^{1 / 2}$ (corresponding to about~$10^{- 35}~{\rm m}$): $D < \hbar^{1 / 2}$. In this case, Eq.~\eqref{line3} leads to
\begin{equation}
    R_\text{S} > D \gg L_\text{clock},
    \label{line4}
\end{equation}
where the last inequality uses that the clock must be much smaller than the rod it measures. Thus, basic QM and general relativity imply that a clock built to measure distances smaller than~$L_\text{P}$ would collapse into a black hole, making such a measurement impossible. 

A similar argument leads to the conclusion that no bona fide clocks able to measure time intervals~$T$ smaller than the Planck time~$T_\text{P} \equiv \hbar^{1 / 2}$ (corresponding to about~$10^{- 43}~{\rm s}$) exist. Let us devote a few lines to illustrate it. (Readers already convinced may skip the rest of this paragraph.) In this case, we shall replace Eq.~\eqref{line2} with
\begin{equation}
    M_\text{cw} \geq \hbar / (2 \delta t) > \hbar / (2 T), 
    \label{line5}
\end{equation}
where the last inequality states that the clock's precision must be enough to measure a time interval~$T$. Combining Eqs.~\eqref{line5} and~\eqref{line1}, we have
\begin{equation}
    R_\text{S} > \hbar / T.
    \label{line6}
\end{equation}
Now, let us consider that~$T < T_\text{P} \equiv \hbar^{1 / 2}$. In this case, Eq.~\eqref{line6} reads
\begin{equation}
    R_\text{S} > T \gg L_\text{clock},
    \label{line7}
\end{equation}
where the last inequality says that a clock must be much smaller than the time interval it measures to be able to process the information; {\em e.g.}, to process a time interval of 1~s, a clock must be much smaller than the Earth-Moon distance. Thus, again basic QM and general relativity imply that a clock built to measure time intervals smaller than~$T_\text{P}$ would collapse into a black hole, making such a measurement impossible. 

The discussion above leads us to the following manifesto: 
\\

\noindent
{\em Given that time is, by definition, what bona fide clocks measure and that there would be no bona fide clocks capable of measuring space and time intervals in the Planck realm, there is a cogent reason to renounce classical spacetimes and theories depending on classical spacetimes in the Planck realm.} 

%%%%%%%%%%%%%%%%%%%%%%%%%%%%%%%%%%%%%%%%%%%%%%%%%%%%%%%%%%
\section{Limits for wave mechanics}
%%%%%%%%%%%%%%%%%%%%%%%%%%%%%%%%%%%%%%%%%%%%%%%%%%%%%%%%%%

\begin{figure}[!ht]
    \begin{center}
    \includegraphics[scale = 0.5]{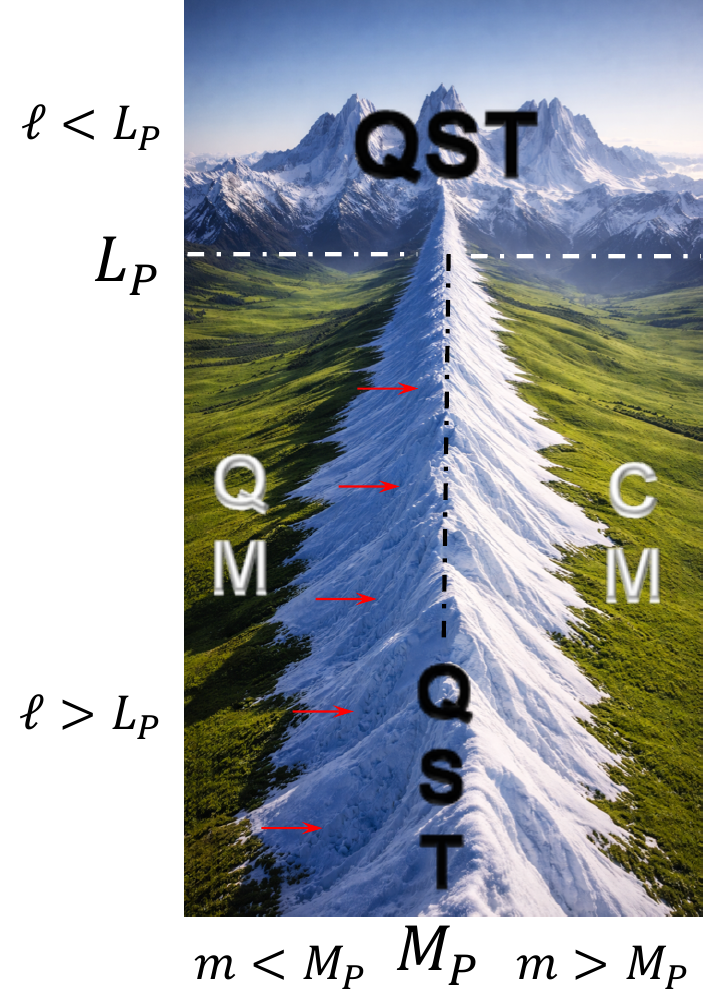}
    \end{center}
    \caption{The figure depicts a landscape representation for our physical scenario. Both QM and CM would emerge on {\em nearly} the same footing from an as yet unknown underlying QST. Even though QM and CM would be emergent theories, it is QM that fixes the Heisenberg cut at~$m \sim M_\text{P} \equiv \hbar^{1 / 2}$. The realm of the QST would be at space scales~$\ell < L_\text{P}$. Still, a QST ridge at~$m \sim M_\text{P}$ would emerge, giving rise to the Heisenberg cut and segregating QM from CM. In this scenario, tabletop experiments aiming to place the c.m. of systems in a spatial superposition should exhibit deviations from QM as their masses approach $M_\text{P}$ due to the QST ridge (red arrows). (Figure generated with AI help.)}
    \label{fig:Estocolmo}
\end{figure}
Quantum field theory is widely regarded as the most precise theory in physics. Quantum electrodynamics matches measurements of the electron’s anomalous magnetic moment to within 1 part in~$10^{12}$. The starting point to construct the Fock space of a field theory is the one-particle Hilbert space given by the Dirac and Klein-Gordon equations,
$$
    (i \slashed{\partial} - \lambdabar^{- 1}) \psi = 0 
    \quad \text{and} \quad
    (\Box + \lambdabar^{- 2}) \phi = 0,
$$
respectively. Both are entirely determined by the (reduced) Compton wavelength~$\lambdabar \equiv \hbar / m c$. It happens that~$\lambdabar$ for fields with masses~$m \gtrsim M_\text{P}$ is~$\lambdabar \lesssim L_\text{P}$. Thus, according to the manifesto above, there is no reason to trust the Dirac and Klein-Gordon equations to rule the c.m. of free systems with masses~$m \gtrsim M_\text{P} \sim 10^{- 5}~{\rm g}$. This calls for a Heisenberg cut to constrain the validity of standard quantum mechanics to fields with~$m \ll M_\text{P}$. 
 
The resulting physical picture is as follows. Standard quantum and classical mechanics~(CM) would be effective emergent theories of a fundamental underlying theory named here (in the absence of a better name) ``quantum spacetime theory''~(QST). (We stress that it is questionable in what sense it would be a ``{\em spacetime theory},'' given the lack of bona fide clocks in its realm.) These two well-mapped regions would describe quantum and classical particles with masses~$m \ll M_\text{P}$ and $m \gg M_\text{P}$, respectively, and would be separated by a QST ridge ruled by new physics. As a result, new physics would be expected not only at scales~$\ell \lesssim L_\text{P} \sim 10^{- 35}~{\rm m}$ but also for particles with masses~$m \sim M_\text{P} \sim 10^{- 5}~{\rm g}$ --- see Fig.~\ref{fig:Estocolmo} for a landscape representation of this physical scenario.
  
%%%%%%%%%%%%%%%%%%%%%%%%%%%%%%%%%%%%%%%%%%%%%%%%%%%%%%%%%%
\section{Quest for the new physics at the Heisenberg cut}
%%%%%%%%%%%%%%%%%%%%%%%%%%%%%%%%%%%%%%%%%%%%%%%%%%%%%%%%%%

In the current context, the ultimate goal is to unveil the QST. Nevertheless, proposing a complete theory in which even the most basic concepts, such as space and time, are likely to be meaningless, sounds as fantastical in the absence of Planck-scale experimental clues as does the discovery of QM, relying only on human imagination and CM. Despite this, we can speculate about how to effectively amend QM to describe systems with masses~$m \lesssim M_\text{P}$. For this purpose, it seems reasonable to assume that it must comply with some minimum requirements:
\begin{enumerate}
    \item Not challenge QM for free particles with~$m \ll M_\text{P}$.
    \item Render free particles with~$m \sim M_\text{P}$ to self-decohere.
    \item Respect Lorentz invariance, meaning that free particles must self-decohere the same in any inertial frames.
    \item Involve no extra scales but the Planck scale, since free particles do not interact with anything (but ``spacetime''). 
\end{enumerate}
Besides item 2 (obviously), note that items 3 and 4 are at odds with the Caldeira–Leggett decoherence mechanism, whose environmental bath privileges a reference frame and requires additional parameters, such as temperature, for its characterization. In Ref.~\cite{AM2025a}, we present a simple gravitational self-decoherence model, satisfying the conditions above. The model proves efficient at decohering particles with~$m \sim M_\text{P}$ (preserving quantum coherence for particles with~$m \ll M_\text{P}$). Such a self-decoherence is to be interpreted as representing a (Lorentz-invariant) leak of quantum information from the particle to (non-observable) quantum degrees of freedom of the QST. In this vein, in contrast to (quark-composed) protons and neutrons, we do not expect, {\em e.g.}, crystals with masses~$m \gg M_\text{P}$ to exhibit interference fringes in double-slit-like experiments (even in ideal conditions). Recently, interference fringes for the center of mass of a cluster of about 7,000 Na atoms, with a total mass of $\sim 170,000~{\rm Da} \sim 10^{- 14}~M_\text{P}$, have been observed in a double-slit-like experiment~\cite{2026-pedalino}. This is still a long way from approaching the Planck mass, but considering that 25 years ago the mass record was set by the fullerene~${\rm C}_{60}$ ($\sim 720~{\rm Da}$)~\cite{1999-arndt}, there is some room for optimism.

In spite of all discussion above, we emphasize that the fact that our scenario rules out the c.m. of systems with~$m \gg M_\text{P}$ from being in spatial superposition, it does not impede such systems from hosting macroscopic {\em (low-energy)} quantum phenomena (see Ref.~\cite{AM2025b} in conjunction with~\cite{AM2025a} for a discussion on it), as those championed by the honoree of this issue of the Brazilian Journal of Physics, Amir Caldeira~\cite{CL81}. In fact, seven Nobel Prizes in Physics (1996, 1998, 2001, 2003, 2007, 2016, and 2025) were awarded for works on this sort of phenomena (see, {\em e.g.}, Refs.~\cite{1972-osheroff_1, 1972-osheroff_2, 1982-tsui, 1983-laughlin, 1995-anderson, 1995-davis, 1950-ginzburg, 1957-abrikosov, 1975-leggett, 1988-baibich, 1989-binasch, 1973-kosterlitz, 1982-thouless, 1983-haldane, 1984-devoret, 1985-martinis, 1985-devoret}).

%%%%%%%%%%%%%%%%%%%%%%%%%%%%%%%%%%%%%%%%%%%%%%%%%%%%%%%%%%
\section{Is nature quantum? Some thoughts and conclusions}
%%%%%%%%%%%%%%%%%%%%%%%%%%%%%%%%%%%%%%%%%%%%%%%%%%%%%%%%%%

Physics does not (or should not) aim at saying “what nature is”; this is an issue for metaphysics. The task of physics is to codify nature’s regularities in terms of mathematical equations. By this token, nature is neither quantum nor classical; {\em nature is what it is}, and there should be no clash in saying that some phenomena obey classical rules, while others obey quantum ones, {\em provided the boundary between these domains is well defined}. This is where the present paper aims to contribute.  

Perhaps the most paradigmatic example of this point of view concerns black holes: {\em are black holes quantum?} Black holes are vacuum solutions of the {\em classical} Einstein's equations. Nevertheless, quantum orthodoxy poses no impediment to black holes satisfying wave mechanics as elementary particles do. By this token, black holes could be placed in spatial superposition, and so on. On the other hand, a black hole of mass~$M_\text{bh}$ has a Schwarzschild radius of~$R_\text{S} = 2 M_\text{bh}$, and its minimum mass should satisfy
$M_\text{bh} \gtrsim M_\text{P} / \sqrt{2}$; otherwise, 
$$
\lambdabar_\text{bh} 
\equiv  \frac{\hbar}{M_\text{bh}} 
\gtrsim \frac{\sqrt{2} \hbar}{M_\text{P}} 
= \sqrt{2} M_\text{P}
\gtrsim 2 M_\text{bh}
= R_\text{S}
$$
and not only would the black hole `leak outside its horizon,' so to speak, but also there would be no bona fide clocks to test them according to the reasoning above, since $R_\text{S} \lesssim L_\text{P}$ (making them scientifically nonsense). The fact that the minimum mass of black holes should satisfy~$M_\text{bh} \gtrsim M_\text{P} / \sqrt{2}$ in conjunction with our Heisenberg cut at~$m \sim M_\text{P}$ would imply that black holes are classical geometrical structures (which is in line with our starting point that black holes are classical solutions of Einstein's equations). Still, perturbations in the vicinity of black holes would evolve quantum mechanically. 

We know that any perturbations disturbing black holes are radiated away as gravitational waves until the black hole relaxes back to a stationary state. Asymptotically at the radiation zone, metric perturbations, $h_{\mu \nu}$, evolve according to the equation  
$$  
\Box \bar{h}_{\mu \nu} = 0 \quad (\bar{h}_{\mu \nu} \equiv h_{\mu \nu} - h^\alpha_{\; \alpha}\eta_{\mu \nu}/2) 
$$ 
in the de Donder gauge, $\nabla_\mu \bar{h}^\mu_{\; \nu}=0$, where $\eta_{\mu \nu}$ is the Minkowski metric. This raises the question: is there {\em any reason whatsoever} to believe that spin-2 massless (graviton) fields could not be quantized, in contrast to spin-1 massless (photon) fields of quantum electrodynamics (the most precise theory ever)? We do not think so, provided the gravitons' energy is small in comparison to the Planck energy (as measured by static observers with respect to the hole). Hence, although black holes would be classical structures according to the proposed Heisenberg-cut scenario, their perturbations should prove possible to quantize.

In summary, we may be looking at fundamental physics from the wrong perspective. Rather than attempting to describe everything, from photons to black holes, using standard quantum mechanics, we should aim at identifying the domains of validity of our standard quantum and classical theories and seek new physical laws at their boundaries. 

% -----------------------
% Back matter
% -----------------------

\backmatter

\section*{Acknowledgements}

The authors are grateful to Atsushi Higuchi and the referees for useful comments. G.E.A.M. is grateful to the organizers of the Fest in honor of Amir Caldeira for the cozy scientific environment. Amir, thank you for all you have done for the Brazilian School of Physics over all these years --- {\em ``live long and prosper''}! 
% BJP requests “Statements and Declarations” (Springer standard)

\section*{Statements and Declarations}

\subsection{Competing interests}
The authors declare no competing interests.

\subsection{Funding}
G.H.S.A. was fully supported by the S{\~a}o Paulo Research Foundation~(FAPESP) under grant~2022/08424-3. G.E.A.M. was partially supported by the National Council for Scientific and Technological Development under grant~301508/2022-4.

\subsection{Data availability}
Not applicable.

\subsection{Code availability}
Not applicable.

% -----------------------
% References
% -----------------------

\end{document}